\documentclass[aaspp4,11pt,preprint,apjfonts]{emulateapj}
 
\slugcomment{Accepted for publication in ApJ}

\shorttitle{Balmer/4000 \AA\ Breaks of Red $z>2$ Galaxies}
\shortauthors{Kriek et al.}
\newcommand{\ha}{H$\alpha$}
\newcommand{\av}{$A_V$}
\newcommand{\wha}{W$_{\rm H\alpha}$}
\newcommand{\td}{$\rm \tau_{320\,Myr}$}

\begin{document}

\title{Direct measurements of the stellar continua and Balmer/4000 \AA\ breaks of red $z>2$ galaxies: Redshifts and improved constraints on stellar populations\altaffilmark{1,2,3}}

\author{Mariska Kriek\altaffilmark{4}, 
Pieter G. van Dokkum\altaffilmark{5}, 
Marijn Franx\altaffilmark{4},  
Natascha M. F\"orster Schreiber\altaffilmark{6}, 
Eric Gawiser\altaffilmark{5},
Garth D. Illingworth\altaffilmark{7}, 
Ivo Labb\'e\altaffilmark{8,9},
Danilo Marchesini\altaffilmark{5}, 
Ryan Quadri\altaffilmark{5}, 
Hans-Walter Rix\altaffilmark{10},
Gregory Rudnick\altaffilmark{11,12}, 
Sune Toft\altaffilmark{5},
Paul van der Werf\altaffilmark{4}, 
and Stijn Wuyts\altaffilmark{4}}

\email{mariska@strw.leidenuniv.nl}

\altaffiltext{1}{Based on observations obtained at the Gemini
Observatory, which is operated by the Association of Universities for
Research in Astronomy, Inc., under a cooperative agreement with the
NSF on behalf of the Gemini partnership: the National Science
Foundation (United States), the Particle Physics and Astronomy
Research Council (United Kingdom), the National Research Council
(Canada), CONICYT (Chile), the Australian Research Council
(Australia), CNPq (Brazil) and CONICET (Argentina).}

\altaffiltext{2}{Based on observations at the W.M. Keck Observatory,
which is operated jointly by the California Institute of Technology
and the University of California.}

\altaffiltext{3}{Based on observations collected at the
European Southern Observatory, Paranal, Chile} 

\altaffiltext{4}{Leiden Observatory, PO Box 9513, 2300 RA Leiden, The
Netherlands}

\altaffiltext{5}{Department of Astronomy, Yale University, P.O. Box
208101, New Haven, CT 06520-8101}

\altaffiltext{6}{Max-Planck-Institut f\"ur extraterrestrische Physik,
Giessenbachstrasse, Postfach 1312, D-85748 Garching, Germany}

\altaffiltext{7}{UCO/Lick Observatory, University of California, Santa
Cruz, CA 95064}

\altaffiltext{8}{Carnegie Observatories, 813 Santa Barbara Street,
Pasadena, CA 91101}

\altaffiltext{9}{Carnegie Fellow}

\altaffiltext{10}{Max-Planck-Institute f\"ur Astronomie, K\"onigstuhl
17, Heidelberg, Germany}

\altaffiltext{11}{National Optical Astronomy Observatory, 950 North
Cherry Avenue, Tucson, AZ 85719}

\altaffiltext{12}{Goldberg fellow}

\begin{abstract} 

We use near-infrared (NIR) spectroscopy obtained with GNIRS on Gemini,
NIRSPEC on KECK, and ISAAC on the VLT to study the rest-frame optical
continua of three `Distant Red Galaxies' (having $J_s - K_s > 2.3$) at
$z>2$. All three galaxy spectra show the Balmer/4000 \AA\ break in the
rest-frame optical. The spectra allow us to determine spectroscopic
redshifts from the continuum with an estimated accuracy $\Delta
z/(1+z)~\sim~0.001-0.04$. These redshifts agree well with the emission
line redshifts for the 2 galaxies with \ha\ emission. This technique
is particularly important for galaxies that are faint in the
rest-frame UV, as they are underrepresented in high redshift samples
selected in optical surveys and are too faint for optical
spectroscopy. Furthermore, we use the break, continuum shape, and
equivalent width of \ha\ together with evolutionary synthesis models
to constrain the age, star formation timescale, dust content, stellar
mass and star formation rate of the galaxies. Inclusion of the NIR
spectra in the stellar population fits greatly reduces the range of
possible solutions for stellar population properties. We find that the
stellar populations differ greatly among the three galaxies, ranging
from a young dusty starburst with a small break and strong emission
lines to an evolved galaxy with a strong break and no detected line
emission. The dusty starburst galaxy has an age of 0.3 Gyr and a
stellar mass of 1 $\times10^{11}$~M$_{\odot}$. The spectra of the two
most evolved galaxies imply ages of 1.3-1.4 Gyr and stellar masses of
4$\times10^{11}$~M$_{\odot}$.  The large breaks for the two most
evolved galaxies indicate that active galactic nuclei (AGN) do not
dominate the rest-frame optical continuum emission of these galaxies,
while for the younger starburst a significant contribution from an AGN
cannot be ruled out.  The large range of properties seen in these
galaxies strengthens our previous much more uncertain results from
broadband photometry. Larger samples are required to determine the
relative frequency of dusty starbursts and (nearly) passively evolving
galaxies at $z\sim2.5$.

\end{abstract}


\section{INTRODUCTION}

Recent studies have demonstrated that galaxies at $z>2$ show a large
range in their rest-frame optical colors. Deep near-infrared (NIR)
imaging has allowed the identification of a new class of $z>2$
galaxies, complementary to the Lyman break galaxies
\citep[LBGs,][]{st96a,st96b} found in optical surveys. \cite{fr03}
introduced the simple color criterion $J_s - K_s > 2.3$, to
efficiently and successfully select red $z>2$ galaxies \citep[Distant
Red Galaxies, or DRGs;][]{fo04,vd04}. The red colors can be caused by
evolved stellar populations, dust, or a combination of both. It is
difficult to assess the origin of the red colors from optical-to-NIR
photometry alone \citep[see][]{fo04}. The extension of photometric
studies to the rest-frame NIR wavelength regime helps: using IRAC on
$Spitzer$, \cite{la05} distinguish the old from the dusty system much
more reliably than was previously possible, and find that a
significant part (30\%) of the DRG sample is indeed best described by
old and passively evolving stellar population models. This result is
supported by the $Spitzer$ 24 $\mu$m imaging presented by \cite{we06},
who find that 65\% of the DRGs host dusty star-forming stellar
populations. Similar findings have also been reported by \cite{pa06}
and \cite{re05} based on optical to mid-infrared photometry of the
GOODS-South and North fields

Although our insight in the nature of $z>2$ galaxies has significantly
improved by the recent access to the rest-frame NIR and IR wavelength
regime, broadband photometric studies have their limitations.  First,
as spectral details get lost, the origin of the observed emission is
uncertain. Emission lines may affect the broadband fluxes, and a
possible contribution by an active galactic nucleus (AGN) is difficult
to quantify.  Second, the modeling results suffer from degeneracies
between age, dust and the star formation history, especially when no
spectroscopic redshift is available.

The large samples of confirmed high-redshift galaxies available today
\citep[e.g.][]{st03,va05} give the impression that obtaining
spectroscopic redshifts is relatively easy. The most efficient and
popular technique in obtaining redshifts for large samples of high
redshift galaxies is multi-object spectroscopy at optical
wavelengths. However, as about 75\% of $2<z<3$ galaxies with
$M>10^{11}\rm M_{\odot}$ have $R>25$ \citep{vd06}, most are well
beyond the limits of optical spectroscopy. Thus NIR spectroscopy is
needed to obtain redshifts for these UV-faint galaxies.

For UV-faint passively evolving galaxies, which lack emission lines
in their rest-frame optical, we are limited to the rest-frame optical
stellar continuum to measure a redshift. The most easily detectable
feature in the spectrum of evolved stars, and thus the best option to
confirm redshifts of $z>2$ evolved galaxies, is the Balmer/4000 \AA\
break. The break is also a powerful diagnostic in stellar population
studies \citep[e.g.][]{ha85,ba99,ka03a}.

For galaxies beyond $z\sim 1.5$ the break shifts into the NIR, where
the combination of high sky background and strong atmospheric
absorption bands complicates continuum studies of galaxies. Recently
new NIR spectrographs on 8-10 meter class telescopes have improved our
access to the NIR regime significantly. This is demonstrated by the
first possible detections of rest-frame optical breaks by \cite{vd04}
and \cite{si04} with KECK/NIRSPEC and SUBARU/CISCO
respectively. Furthermore, the large instantaneous wavelength coverage
offered by GNIRS on Gemini-South \citep[see e.g.][]{vd05} allows, for
the first time, systematic continuum studies of galaxies at $z>2$.

In this paper we present and study the rest-frame optical continua of
three DRGs, of which only one had a spectroscopic redshift prior to
the observations. The data are presented in \S2. In \S3 we fit the
spectra with evolutionary synthesis models and compare the spectra
with the broadband SEDs. In \S4 we use spectral diagnostics to obtain
stellar population properties. We present a direct comparison with
other high-redshift and low-redshift galaxies in \S5. We end with a
summary and conclusions in \S6. Throughout the paper we assume a
$\Lambda$CDM cosmology with $\Omega_{\rm m}=0.3$, $\Omega_{\rm
\Lambda}=0.7$, and $H_{\rm 0}=70$~km s$^{-1}$ Mpc$^{-1}$. All
broadband magnitudes are given in the Vega-based photometric system.

\section{DATA}

\begin{deluxetable*}{crrrllcrrr}
  \tabletypesize{\scriptsize} 
  \tablecaption{Observations\label{obs}}
  \tablewidth{0pt}
  \tablehead{ \colhead{id} & \colhead{$K_s$} & \colhead{$V$} & 
    \colhead{$J_s - K_s$} &
    \colhead{Instrument} &\colhead{date} & \colhead{$\lambda_{\rm
    range}$} & \colhead{Exp(min)} & \colhead{R$_{\rm spec}$} & 
    \colhead{Seeing(\arcsec)}}
  \startdata
    MS1054-1319 & 19.01 & 24.21 & 2.58 & KECK/NIRSPEC & 2003/01/22 & 
    2.05-2.47 & 90 & 1600 & 0\farcs9 \\ 
    & & & & & & 1.28-1.56 & 60 & 1600 & 0\farcs9 \\ 
    & & & & & 2004/02/11 & 1.23-1.55 & 105 & 1600 & 0\farcs8\\  
    & & & & & & 1.50-1.74 & 60 & 1600 & 0\farcs8\\ 
    & & & & VLT/ISAAC & 2004/04/25 & 1.82-2.50 & 81 & 600 & 0\farcs7\\
    & & & & & & 2.14-2.26 & 165 & 3900 & 0\farcs7\\ 
    CDFS-695 & 19.12 & 23.71 & 2.33 & GEMINI-S/GNIRS & 2004/09/02-03 & 
    1.0-2.4 & 92 & 1000 & 0\farcs7\\
    HDFS-5710  & 19.31 & 25.53 & 2.47  & GEMINI-S/GNIRS & 2004/09/06 & 1.0-2.4 
    & 210 & 1000 & 1\farcs0\\ 
  \enddata
\end{deluxetable*}

  \subsection{Target selection and photometry}

  The three studied galaxies lie in the field of MS1054-03, the
  CDF-South and the extended HDF-South respectively. All fields have
  deep NIR imaging and accurate photometry in several optical-to-NIR
  bands. The MS1054-03 field is observed as part of the ``ultradeep''
  Faint InfraRed Extragalactic Survey (FIRES) and the photometry is
  described by \cite{fo06}. The Great Observatories Origins Deep
  Survey \citep[GOODS;][]{gi04} provides deep data of the
  CDF-South. The optical-to-NIR photometry that we used as part of
  this work will be described by Wuyts et al. (2006, in prep.). The
  photometry of the extended HDF-South is part of the new
  MUlti-wavelength Survey by Yale-Chile \citep[MUSYC;][Quadri et
  al. in prep]{ga05}. 

  The three galaxies are chosen on the basis of their $K$ magnitude,
  red $J_s-K_s$ color \citep[$J_s-K_s>2.3$;][]{fr03}, and redshift
  between 2.1 and 2.7. To avoid a bias towards galaxies with strong UV
  emission, we did not require that a previously measured
  spectroscopic redshift was available, and two out of three galaxies
  presented in this paper were selected on basis of their photometric
  redshift. MS1054-1319 was the only galaxy which had a spectroscopic
  redshift ($z=2.423$) prior to these NIR observations, derived from
  both the Ly$\alpha$ and \ha\ emission lines \citep{vd03,vd04}. The
  NIR spectrum of CDFS-695 studied in this paper has already been
  presented by \cite{vd05}, who give an emission line redshift of
  2.225.

  Where possible the broadband fluxes were corrected for emission line
  fluxes. The observed infrared photometry for CDFS-695 is adjusted
  using the line measurements presented by \cite{vd05}. The emission
  line corrections are crucial to interpret the broadband photometry
  for this galaxy, as they account for 0.05, 0.17 and 0.25 mag of the
  total $J$, $H$ and $K$ broadband magnitudes respectively. Hence,
  after corrections CDFS-695 does not satisfy the $J_s-K_s > 2.3$
  criterion. For this galaxy we have no observed optical
  spectrum. Fortunately, CDFS-695 is also part of the {\it K20
  survey}, and from the spectrum presented by \cite{da04a} we conclude
  that the emission lines barely contribute to the observed optical
  broadband fluxes. Thus, no corrections are made to the broadband
  observed optical magnitudes of CDFS-695, but the photometric errors
  are increased by 0.05 mag. For MS1054-1319 the emission line
  correction were derived by \cite{vd04}, who find contributions of
  0.1, 0.02, 0.03, 0.04 mag to the broadband fluxes of $B$, $J$, $H$
  and $K$ respectively, and an increase of the photometric error of
  $V$ and $I$ by 0.05 mag. For HDFS-5710, no emission lines are
  detected in the NIR spectrum and therefore no correction is applied.

  \subsection{Observations} 

  Table 1 summarizes the observations. Two out of three galaxies
  presented in this paper were observed with GNIRS on Gemini-South in
  September 2004: CDFS-695 and HDFS-5710 (program GS-2004B-Q-38). We
  used GNIRS in cross-dispersed mode, in combination with the short
  wavelength camera with the 32 l/mm grating (R=1000) and the
  0\farcs675 by 6\farcs2 slit. In this configuration we obtained a
  wavelength coverage of 1.0 -- 2.4 $\mu m$, divided over 6
  orders. Conditions on September 2 and 3, during which we observed
  CDFS-695, were relatively good with a seeing of 0\farcs7 in $K$ and
  mostly clear sky. HDFS-5710 was observed on September 6 through
  intermittent clouds, and the seeing varied from 0\farcs7 --
  1\farcs3.

  We observed the galaxies following an ABA$'$B$'$ on-source dither
  pattern. In this way we can use the average of the previous and
  successive exposures as the sky frame. The offset between A and A$'$
  ensures that different parts of the detector are used for sky
  subtraction. To optimize the signal-to-noise ratio (S/N) of the
  extracted one-dimensional (1D) spectrum, the shift between A and
  A$'$ should be larger than the area over which the spectrum is
  extracted. Compared to the standard ABBA pattern, this method
  decreases the noise by about 13\,\% in the sky-subtracted frames.

  We complemented the GNIRS spectra with the spectrum of MS1054-1319
  obtained with NIRSPEC on Keck \citep{mc98}. We observed MS1054-1319
  during two runs using the N4 filter and medium dispersion mode
  (1.23-1.55 $\mu$m, R=1600), which is especially useful to target the
  region around the optical break for $z\sim 2.5$ galaxies. The first
  run in January 2003 was characterized by cloudy weather and a
  typical seeing of 0\farcs9 in $K$ \citep{vd04}. The conditions
  during the second run in February 2004 were somewhat better with a
  seeing of 0\farcs8.  To extend the wavelength coverage for this
  galaxy, we complement the N4 coverage by $H$ and $K$ band spectra
  obtained with NIRSPEC on KECK and ISAAC on the VLT. For details on
  these observations see Table~\ref{obs}. The NIRSPEC and ISAAC
  spectra were taken at three dither positions (ABC), as the slit
  length of both instruments is longer than for GNIRS. Similar to the
  ABA$'$B$'$ pattern we can use the average of the previous and
  following exposures as sky frame.

  All targets were acquired by blind offsets from nearby stars. With
  NIRSPEC the alignment was checked and corrected if needed before
  every individual 900\,s exposure. Furthermore, at each dither
  position a separate spectrum was taken of the offset star to
  determine the expected position of the object spectrum on the
  detector. As GNIRS and ISAAC have better pointing stability there
  was no need for re-acquisition after each individual exposure, and
  acquisition checks were performed approximately every hour. The
  total exposure times of all objects and modes are listed in
  Table~\ref{obs}.

  After each observing sequence, we observed an AV0 star near the
  target. These spectra are used in the reduction to correct for
  detector response and atmospheric absorption.

  We note that other targets were also observed during these
  runs. However, the galaxies discussed here are the only ones for
  which we specifically attempted to measure the continuum rather than
  just emission lines.

    \begin{figure*}
    \begin{center}
      \epsscale{1.1}
      \plotone{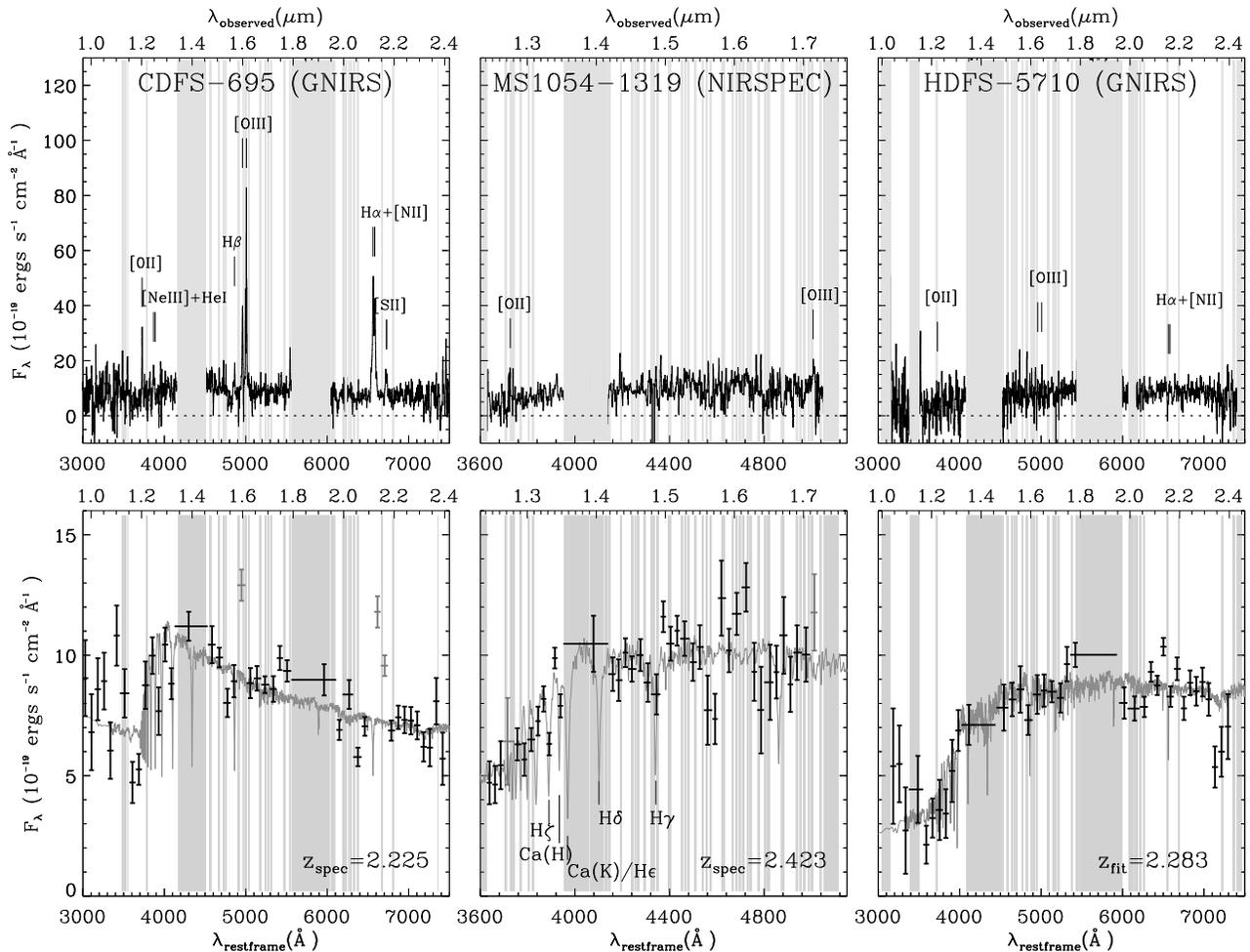}
      \caption{Extracted one-dimensional spectra ({\em upper panels})
	and their binned version ({\em lower panels}) for CDFS-695
	({\em left}), MS1054-1319 ({\em middle}) and HDFS-5710 ({\em
	right}).  The presented original 1D spectra (upper panels) are
	smoothed by a boxcar of 25 \AA\ and 13 \AA\ (observed frame)
	for GNIRS and NIRSPEC spectra respectively. Bad atmospheric
	regions or wavelengths that are heavily contaminated by
	skylines are indicated in {\it light grey}. Bins that include
	emission line fluxes are plotted in {\it grey} as well. The
	best fit template spectra ({\it dark grey}) are overplotted
	(see Table~\ref{best}). For each galaxy a break is detected
	between 3650 and 4000 \AA.
	\label{spec}}
    \end{center}
\end{figure*}

  \subsection{Reduction of GNIRS spectra}

  The reduction of the NIRSPEC spectra of MS1054-1319 is described by
  \cite{vd04}. The ISAAC spectra of MS1054-1319 were reduced in a
  similar way as the NIRSPEC spectra. For the GNIRS data reduction of
  CDFS-695 and HDFS-5710 we developed a suite of custom scripts to
  perform the following steps. We started the reduction by removing a
  bias pattern. The pattern repeats itself every 8 columns,
  differently in each quadrant. It was modeled by fitting a constant
  along the vertical direction in regions where no spectrum is
  present, and removed over the whole detector. We also corrected for
  persistence from the acquisition image on the detector.

  Cosmic rays were removed as follows. First, we removed the sky from
  the science frames by subtracting the average of the previous and
  next exposure. For additional sky-subtraction we have to straighten
  the orders, which follow curved paths on the detector. We determined
  the position of the object spectrum in each row of each order by
  tracing the spectrum of a bright star. Next, we straightened the
  orders by integer pixel shifts (to retain the cosmic ray shapes),
  using the expected object positions. Because of the small slit
  length, there is not enough `empty' sky left to remove the remaining
  sky by fitting lines along the spatial direction. However, the
  negative object spectrum should cancel its positive counterpart, and
  any residual sky was removed by requiring that the biweight mean
  \citep{be90} along the spatial direction is zero at every
  wavelength. Next, we run L.A.Cosmic \citep{vd01} on the sky-free
  frames to identify the cosmic rays.  The resulting cosmic ray masks
  were convolved by a boxcar to exclude neighboring pixels. The masks
  of the different orders were transformed back to the original shape
  of the spectrum without interpolation and combined to one cosmic ray
  mask for each individual exposure. Throughout the following
  description, the cosmic ray mask will be transformed in the same way
  as the science frames.

  We continued with the original raw frames and performed the same
  steps as described above to remove the sky, only this time we
  allowed interpolation when straightening the orders. For each
  exposure and each order, a 1D noise frame was made from the
  subtracted sky. This noise frame was used to identify dead and hot
  pixels, which were added to the mask.

  Next, the orders of each frame were wavelength calibrated using arc
  lamp frames. The calibrations were checked and if necessary corrected
  using skylines. Corrections for the instrumental response function
  and atmospheric absorption were applied by dividing by the response
  spectrum.  This spectrum was created from the observed spectrum of an
  AV0 star, which was divided by the spectrum of Vega. Residuals from
  Balmer absorption features in the spectrum of the AV0 star were
  removed by interpolation. The spectrum of the AV0 star was reduced in
  the same way as the science frames.

  The different exposures were combined for each order, excluding
  cosmic rays and outliers as identified in the masks. Finally the
  orders were combined, properly weighting overlapping regions using
  the response function. In the entire procedure pixels were
  interpolated only once, to minimize smoothing and noise correlations
  in the final frames.

  \subsection{Extraction of 1 dimensional spectra}\label{extr}

  For all three galaxies the 1D spectrum was extracted by summing all
  lines (along the wavelength direction) with a mean flux $>$ 0.25
  $\times$ the flux in the central row, using optimal weighting. In
  addition, a `low-resolution' (binned) spectrum was extracted as
  follows. The pixels were sampled along the wavelength direction in
  each line of the two-dimensional (2D) spectrum using the biweight
  estimator. Thus we created a binned spectrum for each line of the 2D
  spectrum. Hereafter the lines with a mean flux $>$ 0.25 $\times$ the
  flux in the central line were added corresponding to their weighting
  factors. This procedure gives a higher S/N in the final low
  resolution spectrum than binning the original extracted 1D spectrum.

  Regions with low or variable atmospheric transmission or with strong
  sky line emission were excluded.  For CDFS-695 and HDFS-5710
  observed with GNIRS, we defined these regions as those with less
  than 5\% of the maximum transmission and with sky line intensities
  of 30\% or more of the strongest line. For MS1054-1319 observed with
  NIRSPEC and ISAAC, these criteria were $<$30\% and $>$25\%,
  respectively. For MS1054-1319 the spectra were sampled in 30 pixel
  bins ($\sim$80 \AA), and the other two galaxies were sampled in bins
  of 50 pixels ($\sim$250 \AA). Rather than to vary the number of
  pixels that contribute to each wavelength bin (i.e. excluding bad
  wavelength regions), the width was adjusted such that each bin
  contains 30 or 50 `good' pixels.

  For MS1054-1319 we combined the NIRSPEC and ISAAC spectra of
  different but overlapping wavelength coverages. Adding 2D spectra
  would have decreased the S/N as there is a difference in seeing and
  sampling of instruments. Instead we combined the different spectra
  in the following way. First, we extracted a low resolution spectrum
  for each individual 2D spectrum, by using the same bins and a common
  mask for excluding regions of poor atmospheric transmission or with
  strong sky lines. These spectra were scaled by minimizing $\chi^2$
  in the overlapping regions and thereafter combined. These scaling
  factors were used to combine the `high resolution' 1D spectra as
  well. Spectra of different wavelength coverages without overlapping
  regions were scaled using the total broadband $H$ and $K$
  magnitudes. To scale the spectra, we modeled the continuum by
  fitting a straight line to the low resolution spectrum, excluding
  bins that may contain emission lines. The modeled continuum was
  converted to $F_{\nu}$ and thereafter convolved with the appropriate
  filter-curve and normalized using the emission line corrected total
  fluxes. The GNIRS spectra of HDFS-5710 and CDFS-695 were scaled in
  the same way, but as the wavelength coverage is not divided over
  different bands, we used an error-weighted average scale factor
  obtained from the broadband $H$ and $K$ magnitudes to tie the
  spectrum to the broad band photometry. In this procedure, we did not
  use the broadband $J$ magnitude because the Balmer/4000 \AA\ break
  complicates the modeling of the continuum in this band.

  The 1D original and low resolution spectra are presented in
  Fig.~\ref{spec}.  For MS1054-1319 we only show the area around the
  break observed with NIRSPEC, which has continuous coverage. For each
  galaxy an optical break is detected. This is no coincidence, as the
  selection criterion of $J_s-K_s > 2.3$ was introduced to identify
  galaxies with prominent 4000~\AA~ and/or Balmer breaks. There are
  even indications that some individual absorption features are
  detected for MS1054-1319 (H$\zeta$ and Ca(H)) and CDFS-695 (Ca and
  H$\delta$). The sensitivity of the spectra of CDFS-695 and
  HDFS-5710, observed with GNIRS, decreases in the blue, which is
  reflected in the error bars in Fig.~\ref{spec}. The outliers
  bluewards of the break in CDFS-695 are caused by residual bias
  patterns which cannot be properly removed.

  In the spectrum of CDFS-695, we clearly detect several emission
  lines that are labeled in Fig.~\ref{spec}. The analysis of these
  emission lines is presented by \cite{vd05}.  For MS1054-1319 we
  observe two possible emission lines ([0\,{\sc II}] at 3727 \AA\ and
  [0\,{\sc III}] at 5007 \AA) in the combined $N4$ and $H$ band
  spectra. This galaxy also has detected H$\alpha$ and [N\,{\sc ii}]
  emission in its K-band spectrum \citep{vd04}. Strikingly, for
  HDFS-5710 no emission lines were detected at all. Note that the
  galaxy without emission line detections appears to have the
  strongest break (HDFS-5710), and that the galaxy with the strongest
  lines (CDFS-695) has the weakest break.  In \S~\ref{SFH} we will
  investigate this relation between the break strength and emission
  lines in more detail.

\begin{figure}
  \begin{center}
    \epsscale{1}
    \plotone{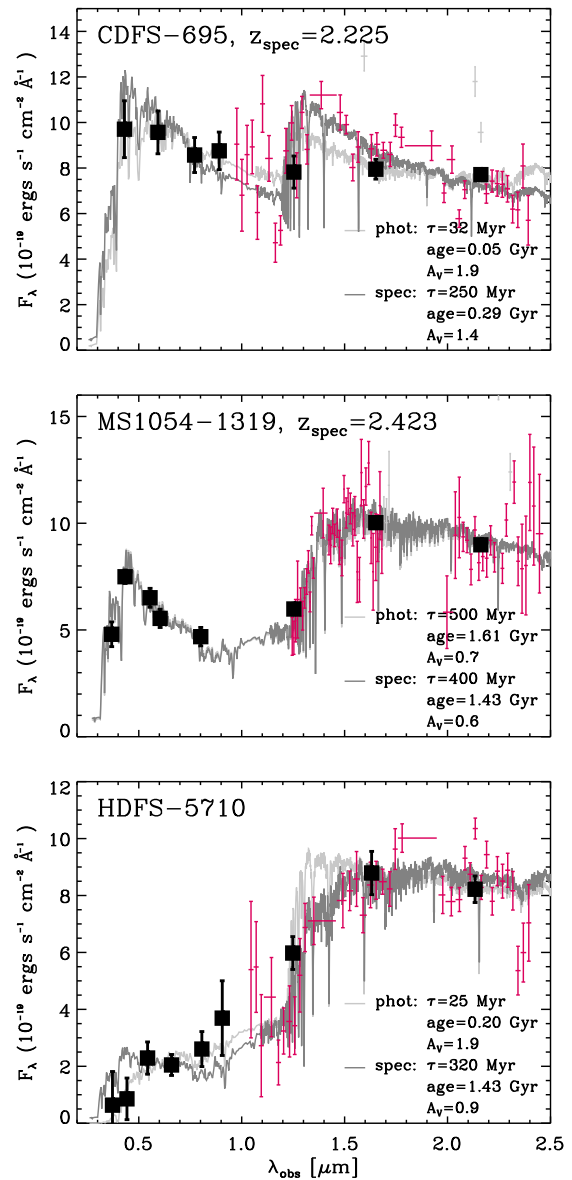}
    \caption{Broadband photometry together with the observed and
      modeled spectra for CDFS-695 ({\it upper panel; B, V, I, z, J,
      H, \& K}), MS1054-1319 ({\it middle panel; U, B, V, R, I, J, H,
      \& K}), and HDFS-5710 ({\it lower panel; U, B, V, R, I, z, J, H,
      \& K}). The error bars represent the 1$\sigma$ uncertainties of
      the flux measurements. The bins that are contaminated by
      emission line fluxes, and thus not included during fitting are
      plotted in grey. The best fits to just the photometry ({\em
      light grey}) and to the spectra in combination with the
      rest-frame UV bands ({\em dark grey}) are overplotted. For
      MS1054-1319 and CDFS-695 the redshift was fixed to the emission
      line redshift during the fitting procedure, while for HDFS-5710
      $z$ was a free parameter. The $\tau$, $A_v$, and $age$ (see
      \S3.1) of the best fit are listed in the panels. \label{fits}}
  \end{center}
\end{figure}

\begin{deluxetable*}{llllcrrrrlll}
  \tabletypesize{\scriptsize} 
  \tablecaption{Modeling Results\label{best}}  
  \tablewidth{0pt}
  \tablehead{ \colhead{id} & \colhead{$\tau$} & \colhead{age} &
  \colhead{$A_V$} & \colhead{$z$\tablenotemark{a}} &
  \colhead{$\chi^{2}_{\rm min}$\tablenotemark{b}} &
  \colhead{$\chi^{2}_{\rm 1\sigma}$} & \colhead{$\chi^{2}_{\rm
  2\sigma}$} & \colhead{N\tablenotemark{c}} & \colhead{$M_{\star}$} &
  \colhead{SFR} & \colhead{SFR/$M_{\star}$} \\ & \colhead{Gyr} & 
  \colhead{Gyr} & \colhead{mag} & & & & & & \colhead{$10^{11}$ M$_{\sun}$} 
  & \colhead{M$_{\sun}$yr$^{-1}$} & \colhead{$10^{-11}$ yr$^{-1}$}}
    \startdata 
    CDFS-695 & & & & & & & & & & & \\

    ~phot & 0.03$^{+9.97}_{-0.02}$ & 0.05$^{+0.15}_{-0.00}$ & 
    1.9$^{+0.2}_{-0.4}$ & - & 0.77 & 1.46 & 3.46 &   3 & 
    0.8$^{+0.3}_{-0.1}$ &  722$^{+ 874}_{- 662}$ &  932$^{+1276}_{- 854}$\\

    & & & & & & & & & & & \\
 
    ~spec & 0.25$^{+9.75}_{-0.22}$ & 0.29$^{+0.28}_{-0.18}$ & 
    1.4$^{+0.1}_{-0.2}$ & - & 2.78 & 2.85 & 2.95 &  
    39 & 1.3$^{+0.3}_{-0.3}$ &  284$^{+ 100}_{- 206}$ & 226$^{+  85}_{- 143}$\\

    & & & & & & & & & & & \\

    \tableline

    MS1054-1319 & & & & & & & & & & & \\    

    ~phot & 0.50$^{+0.50}_{-0.18}$ & 1.61$^{+0.79}_{-0.47}$ &
    0.7$^{+0.2}_{-0.1}$ & - & 1.28 & 2.10 & 3.20 & 4 &
    4.2$^{+1.8}_{-1.3}$ & 47$^{+ 34}_{- 11}$ & 11$^{+ 6}_{- 1}$ \\

    & & & & & & & & & & & \\

    ~spec & 0.40$^{+0.10}_{-0.08}$ & 1.28$^{+0.16}_{-0.14}$ &
    0.7$^{+0.2}_{-0.1}$ & - & 1.59 & 1.67 & 1.73 & 61 &
    3.5$^{+0.6}_{-0.4}$ & 48$^{+ 25}_{- 12}$ & 14$^{+ 6}_{- 2}$ \\

    & & & & & & & & & & &\\

    \tableline

    HDFS-5710 & & & & & & & & & & &\\    

    ~phot & 0.03$^{+9.98}_{-0.01}$ & 0.20$^{+2.80}_{-0.15}$ &
    1.9$^{+1.1}_{-1.9}$ & 2.28$^{+0.15}_{-1.30}$ & 0.52 & 1.90 &
    2.83 & 4 & 3.2$^{+2.1}_{-3.0}$ & 4$^{+1417}_{- 4}$ & 1$^{+ 546}_{-1}$ \\

    & & & & & & & & & & & \\

    ~spec & 0.32$^{+0.18}_{-0.16}$ & 1.43$^{+1.32}_{-0.93}$ &
    0.9$^{+1.1}_{-0.6}$ & 2.283$^{+0.028}_{-0.020}$ & 2.21 & 2.37 &
    2.57 & 38 & 4.0$^{+1.7}_{-0.8}$ & 19$^{+ 165}_{- 15}$ & 4$^{+33}_{- 3}$ 
    \enddata
      \tablenotetext{a}{The redshifts of CDFS-695 and MS1054-1319
	are fixed to z$_{\rm H\alpha}$=2.225 and z$_{\rm
	  H\alpha}$=2.423 respectively.} 
      \tablenotetext{b}{$\chi^2$ is given per degree of
	freedom. The 1$\sigma$ and 2$\sigma$ values correspond to
	the 68\% and 95\% confidence levels, derived from the
	Monte Carlo simulations described in \S3.1} 
      \tablenotetext{c}{Number of degrees of freedom}
\end{deluxetable*}

\section{SPECTRAL MODELING}\label{mod}

Modeling the broadband photometry is a popular method to study the
properties of high-redshift galaxies
\citep[e.g.][]{sa98,pa01,sh01,fo04}. Unfortunately, the modeling
results suffer from degeneracies between age, dust and the star
formation timescale. In this section we investigate the additional
constraints provided by the rest-frame optical spectra.

  \subsection{Fitting procedure and results}\label{proc}

  We used synthetic spectra by \cite{bc03} to model our data. We
  selected a stellar population with a \cite{sa55} initial mass
  function (IMF) between 0.1-100 M$_{\odot}$, a solar metallicity, and
  based on the Padova 1994 evolutionary tracks. We adopted the
  reddening law of \cite{ca00} and corrected for intergalactic
  attenuation by the Ly$\alpha$ forest using the prescriptions of
  \cite{ma96}. For a discussion about the choice of the model
  parameters and the effects of their variations, see \cite{fo04}.
  
  The star formation history is parametrized by an exponentially
  declining function with characteristic time scale $\tau$. Synthetic
  spectra were constructed for a grid of 31 different $\tau$'s between
  10 Myr and 10 Gyr, 24 different ages between 10 Myr and 3 Gyr (and
  always restricted to be less than the age of the universe at a given
  redshift), redshifts in steps of $\Delta z = 0.001$, and 31
  extinction values (A$_V$) between 0 and 3 mag. For MS1054-1319 and
  CDFS-695, the redshift was fixed to the emission line redshift.

  The templates were sampled in the same bins as the observed spectra,
  thus wavelength regions with bad atmospheric transmission and strong
  skylines in the observed spectra were excluded in the synthetic
  spectra as well. Additionally, bins that were contaminated by
  emission line fluxes ([OIII], \ha, [NII] and [SII]) were
  excluded during fitting, as the models do not include emission by
  the ionized interstellar medium. To extend our wavelength coverage
  to the rest-frame UV, we complemented the observed spectra with the
  optical photometry. The broadband fluxes of the synthetic spectra
  were derived by integrating the flux accounting for the filter
  curves. The minimum error in the broadband fluxes was set to 0.05
  mag to account for absolute calibration uncertainties.

  The spectra in combination with the rest-frame UV photometry were
  fitted by using least-square minimization ($\chi^2$). We performed
  200 Monte Carlo simulations to calibrate the confidence levels. For
  these simulations we varied the fluxes randomly within the errors,
  assuming a Gaussian distribution, and followed the same procedure as
  applied to the original data. The 1$\sigma$ and 2$\sigma$ confidence
  levels are derived from the $\chi^2$ values of the original fit
  which enclose the best 68\% and 95\% of the simulations \citep[see
  ][]{pa01}. These $\chi^2$ values per degree of freedom are listed in
  Table 2.

\begin{figure*}
  \begin{center}
    \epsscale{1.1}
    \plotone{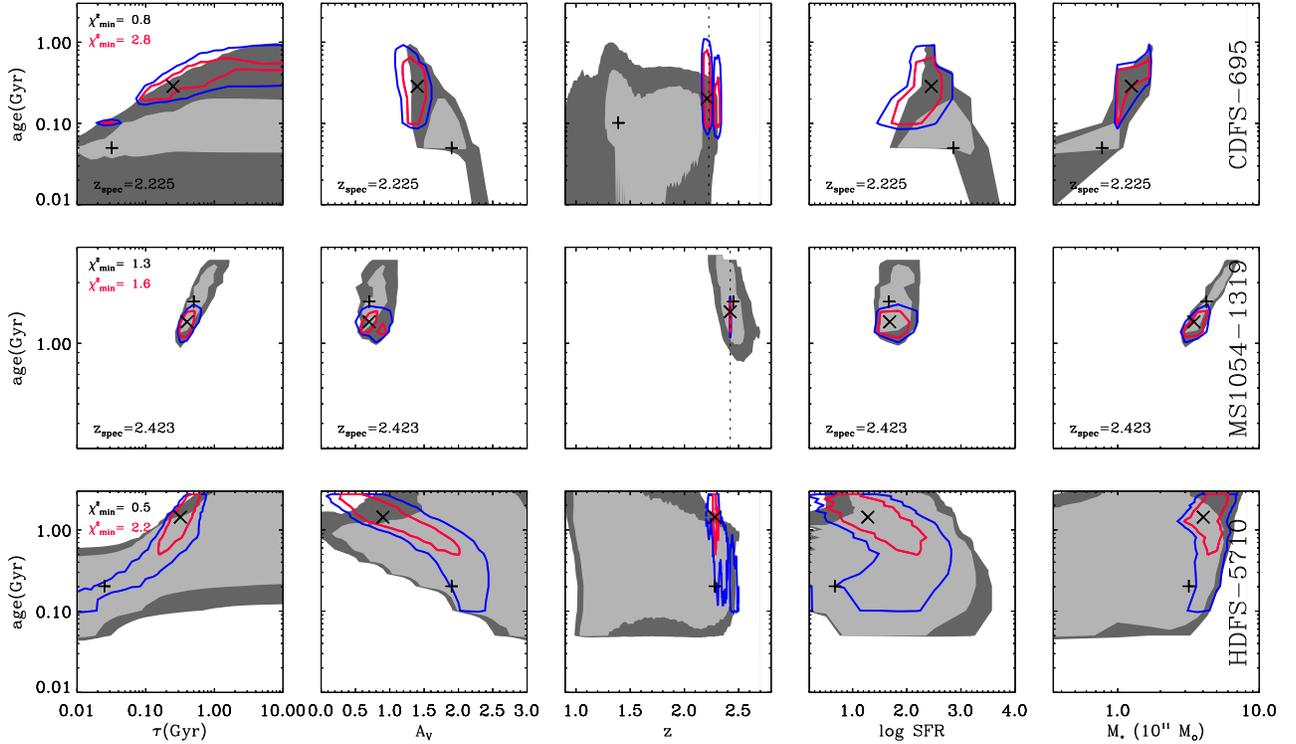}
    \caption{Comparison of modeling results between the spectrum plus
      rest-frame UV broadband fluxes and the broadband SED for
      CDFS-695 ({\em upper panels}), MS1054-1319 ({\em middle panels})
      and HDFS-5710 ({\em lower panels}). The best-fit solutions and
      their 1$\sigma$ and 2$\sigma$ confidence levels are presented by
      the crosses and color contours for the spectrum together with
      the optical photometry, and pluses and filled greyscale contours
      for solely the optical-to-NIR photometry. For MS1054-1319 and
      CDFS-695, the redshift was fixed to $z_{\rm H\alpha}$ to
      derive $\tau$, $age$, $A_V$, SFR and stellar mass. The dotted
      lines in the age versus $z$ plots indicate the emission line
      redshifts. For HDFS-5710 the redshift was a free parameter when
      deriving all properties. The $\chi^2$ value of the best fit
      solution is given in the left panels, in red for the spectral
      fit and in black for the photometry fit. This figure shows that
      the spectrum improves the constraints on all properties
      substantially. In addition, the modeled redshifts for CDFS-695
      and MS1054-1319 are remarkably accurate.\label{conf}}
  \end{center}
\end{figure*}

  The best spectral fits are shown in Fig.~\ref{fits}. The $\tau$,
  age, \av, redshift, stellar mass, star formation rate (SFR) and
  $\chi^2$ corresponding to these fits are listed in
  Table~\ref{best}. All allowed combinations of $z$, $\tau$, \av\ and
  $age$ are presented as 1$\sigma$ and 2$\sigma$ confidence levels
  (color contours) in Fig.~\ref{conf}. 

  \subsection{Continuum redshifts}

\begin{figure}
  \begin{center}
    \epsscale{1.1}
    \plotone{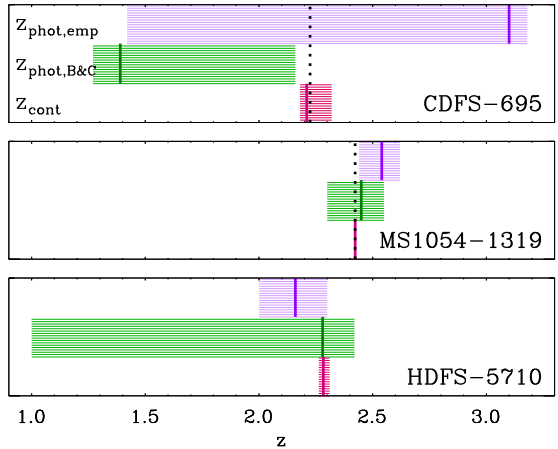} 
    \caption{Comparison between continuum redshift, broadband
      photometric redshift and the emission line redshift. The
      best-fit redshift and its 1$\sigma$ confidence interval to the
      NIR continuum together with the optical photometry is indicated
      by the red line and surrounding shaded area respectively. The
      green thick line and the surrounding shaded area present the
      best-fit redshift and its allowed 1$\sigma$ range when fitting
      the optical-to-NIR broadband photometry by \cite{bc03}
      models. The best-fit redshift and its 1$\sigma$ uncertainty
      obtained when fitting the photometry by linear combinations of
      empirical and theoretical galaxy templates \citep{ru01,ru03} is
      shown by the purple thick lines and shaded areas. The emission
      line redshift, if present, is indicated by the black dotted
      line. This diagram illustrates that the continuum shape is
      powerful in obtaining accurate and well-constrained
      redshifts. \label{redshifts}}
  \end{center}
\end{figure}

  To test the accuracy of the redshift obtained by spectral fitting,
  both CDFS-695 and MS1054-1319 were refit with $z$ as a free
  parameter. The spectra of these galaxies give remarkably good
  solutions for $z$. This is illustrated in Fig.~\ref{redshifts}. For
  MS1054-1319, which has an H$\alpha$ and Ly$\alpha$ redshift of 2.423
  and 2.424 respectively \citep{vd03,vd04}, we find $z_{\rm fit} =
  2.423^{+0.002}_{-0.003}$. For CDFS-695 which has an H$\alpha$ and
  Ly$\alpha$ redshift of 2.225 and 2.223 \citep{vd05,da04a}, we find
  $z_{\rm fit} = 2.21^{+0.11}_{-0.03}$. This galaxy has a secondary
  solution for redshift ($z\sim2.32$) that is probably caused by
  mis-identification of the absorption features in this spectrum.

  The uncertainty on the derived value for the redshift is decreased
  significantly compared to only using the broadband fluxes. The green
  lines and shaded areas in Fig.~\ref{redshifts} indicate the
  1$\sigma$ redshift range allowed by the photometry, using
  \cite{bc03} models (see confidence levels in Fig.~\ref{conf}). The
  purple lines and surrounding shaded areas are obtained by fitting
  linear combinations of theoretical and empirical templates to the
  broadband photometry, following the method described by
  \cite{ru01,ru03}. The uncertainty on the redshifts are decreased by
  a factor of 6-50 when including the spectrum.

  For HDFS-5710 which is the only galaxy without an emission line
  redshift, spectral modeling yields $z_{\rm
  fit}=2.283^{+0.028}_{-0.020}$ (see Fig.~\ref{redshifts}).  This
  result illustrates the potential of NIR spectroscopy to measure
  redshifts of $z>2$ galaxies that are faint in the rest-frame UV and
  have no detectable emission lines.

  \subsection{Comparison to broadband SEDs}\label{prop}

  To investigate if the spectra confirm and improve the constraints on
  the stellar populations properties, we refitted the photometry of
  the three galaxies, ignoring the spectral information.  The
  comparison between the allowed solutions for the input parameters $z,
  \tau, age$ and \av, and the output parameters stellar mass and SFR
  for the spectra and broadband SED is presented in
  Fig.~\ref{conf}. The corresponding values are listed in
  Table~\ref{best}. The best-fit templates to the spectra and
  photometry, and the best fits to the photometry alone are shown in
  Fig.~\ref{fits}.

  Fig.~\ref{conf} shows that the spectra greatly improve the
  constraints on the modeled properties, for all three galaxies. Most
  of the 1$\sigma$ solutions to the spectra fall within the 2$\sigma$
  confidence levels of the photometry. However, most of the 1$\sigma$
  confidence regions of the photometric fits are, within 2$\sigma$,
  not allowed by the spectral fits.  For example, while the photometry
  of CDFS-695 still allows a wide range in dust content and age, the
  solutions dustier than 1.6 mag and younger than 0.1 Gyr are ruled
  out by the spectrum. Even MS1054-1319, for which the properties are
  already well-constrained by the photometry, the spectrum decreases
  the 1$\sigma$ confidence interval of age and $\tau$ by a factor
  7-8.  And finally, for HDFS-5710 the largest gain is achieved by the
  better constrained redshift. Also, solutions with $\tau$ longer than
  700 Myr are ruled out by the spectrum.
 
  The normalized $\chi^2$ values per degree of freedom for the fits to
  the broadband photometry are lower than those for the fits including
  the NIR spectroscopy. However, the difference between the $\chi^2$
  of the best fit and the 1$\sigma$ and 2$\sigma$ confidence intervals
  are smaller for fits including the NIR spectra (see
  Table~\ref{mod}), which result in the tighter constraints. One
  possible cause for the difference in minimal $\chi^2$ of the best
  photometric and spectral fits is that $\chi^2$ statistics are very
  sensitive to the data uncertainties. The photometric errors may be
  overestimated or the spectral errors may be underestimated. Another
  cause could be template mismatch, which can become more problematic
  with more detailed spectroscopic information. 

  To test the robustness of our results we repeated the fitting
  procedure for MS1054-1319, artificially decreasing the errors on the
  photometry by a factor of 2 and increasing the errors on the spectra
  by a factor of 2. The allowed ranges for $A_V$, age and $\tau$ are
  slightly increased and reduced for the spectra and photometry
  respectively, but even in this extreme case inclusion of the spectra
  still improves the constraints on the stellar populations.

  \subsection{Discussion of modeling results}

  Our fitting results show that including the NIR spectra when fitting
  the broadband fluxes sets tighter constraints on stellar population
  properties and reduces degeneracies between age, extinction and the
  SFH.  This is mainly due to the measurement of the shape and
  strength of the optical break. We conclude that the spectral shape
  in combination with the photometry is more powerful than the
  broadband photometry alone in tracing the properties of stellar
  populations in high redshift galaxies.

  The DRGs studied here span a wide range in their derived properties,
  consistent with results of larger samples based on modeling of their
  broadband SEDs \citep[e.g.][]{fo04,la05}. Their SFH, reddening and
  age are quite different, ranging from CDFS-695 which is about 0.2
  Gyr old and is forming a few hundred solar masses a year, to
  HDFS-5710 and MS1054-1319 which are both best fitted by a stellar
  population of 1.4 Gyr old and are forming a few tens of solar masses
  a year. 

  \cite{la05} divided their DRG sample into dusty starburst and ``red
  and dead'' galaxies based on the observed NIR and IR colors ($I-K$
  and $K-4.5\mu$).  To compare the DRGs presented in this work with
  those of \cite{la05}, we measured the same colors by extrapolating
  the best spectral fit to the rest-frame NIR. All three galaxies fall
  in the dusty star-forming part of the plot, although their locations
  move closer towards the ``red and dead'' part, when going from
  CDFS-695 to MS1054-1319 to HDFS-5710.

\section{SPECTRAL DIAGNOSTICS}\label{spec_diag}

In the previous section we studied the spectra by comparing them to
models. The break appeared to be the driving feature behind the
improved modeling constraints. In this section, we take a different
approach by measuring the break directly, and use this feature in
combination with \ha\ to constrain the star formation histories of the
galaxies.

\subsection{Break indices}\label{sp_ind}

  The Balmer and 4000 \AA\ break are often treated as one feature,
  owing to their similar locations and the fact that they partially
  overlap. However, the breaks originate from different physical
  processes, and behave differently as populations age. Both breaks
  are due to absorption in the atmosphere of stars. The 4000 \AA\
  break arises because of an accumulation of absorption lines of
  mainly ionized metals. As the opacity increases with decreasing
  stellar temperature, the 4000 \AA\ break gets larger with older
  ages, and is largest for old and metal-rich stellar populations. The
  metallicity is of minor influence for ages less than 1 Gyr
  \citep{bc03}. There are several definitions introduced to quantify
  the strength of the 4000 \AA\ break. \cite{br83} proposed $D(4000)$,
  which measures the ratio of the average flux density $F_{\nu}$ in
  the bands 4050-4250 \AA\ and 3750-3950 \AA\ around the
  break. Because of the broad regions, this index is fairly sensitive
  to reddening by dust. To reduce this effect, \cite{ba99} defined a
  new index $D_n(4000)$, based on smaller continuum regions (3850-3950
  \AA, 4000-4100 \AA).

  The Balmer break at 3646\,\AA\ marks the termination of the Hydrogen
  Balmer series, and is strongest in A-type stars. Therefore the break
  strength does not monotonically increase with age but reaches a
  maximum in stellar populations of intermediate ages (0.3-1 Gyr). The
  strength of the Balmer sequence can be best measured from the
  individual Balmer lines, such as H$\delta$. However, as our spectra
  do not allow the measurement of this feature, we use the strength of
  the Balmer break ($D_B$), which we define as the ratio of the
  average flux density $F_{\nu}$ in the bands 3500-3650 \AA\ and
  3800-3950 \AA\ around the break. The large regions are not optimal,
  but are a trade-off between dust-dependence and having sufficient
  S/N using the spectra of high redshift galaxies.  This index is also
  partially influenced by the 4000 \AA\ break. Nevertheless, the
  age-dependence of $D_B$ is very similar to that of H$\delta$.

  We determined $D_n(4000)$ and $D_B$ for the three DRGs.  The average
  flux in the regions around the breaks is measured in the same way as
  we extracted the binned spectra. For MS1054-1319 it was impossible
  to measure the flux redwards of the 4000 \AA\ break, as the defined
  region is entirely covered by an atmospheric band. Therefore we
  measured $D(4000)$ and used the \cite{bc03} models to derive
  $D_n(4000)$ for this galaxy. When converting the indices we allowed
  an \av\ between 0 and 3 and all three star formation histories to
  determine the uncertainty. We note that this only slightly increases
  the uncertainty as for the measured break strength of MS1054-1319
  the correlation between $D(4000)$ and $D_n(4000)$ is not very
  dependent on dust or the star formation history.

\begin{figure}
  \begin{center}
    \epsscale{1.1}
    \plotone{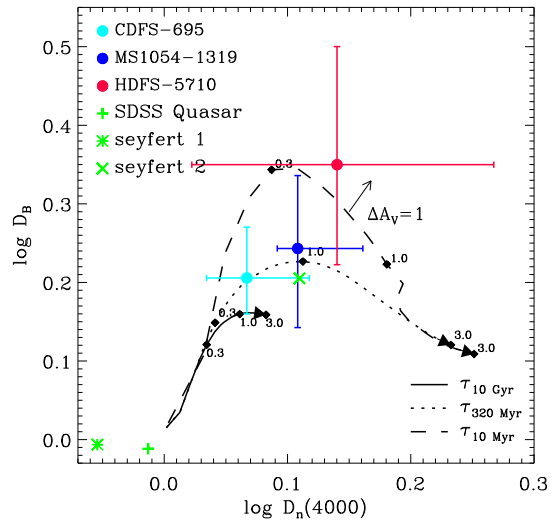} 
    \caption{The Balmer break versus the 4000 \AA\ break for the
      three studied galaxies and three different star-formation
      models. The small black diamonds and the corresponding values
      indicate the age in Gyr.  The effect of attenuation is indicated
      as a vector in the plot. Measurements for three type of AGNs are
      shown as well \citep{fr91}. This figure illustrates that a
      dust-free $\tau_{\rm 10 Gyr}$ (which is comparable to constant
      star formation for galaxies at $z>2$) is unable to produce large
      breaks within 3 Gyr. For MS1054-1319 the combined break implies
      a stellar population of about 1 Gyr, while for CDFS-695 we find
      a younger stellar population. The large Balmer break of
      HDFS-5710 suggests that this galaxy is in a post-starburst
      phase, and that $\tau_{\rm 10 gyr}$ or CSF is highly
      unlikely. The large breaks for MS1054-1319 and in particular
      HDFS-5710 imply that the optical spectrum is dominated by
      stellar light.\label{break_vs_break}}
  \end{center}
\end{figure}

  The break measurements are listed in Table~\ref{diag} and shown in
  Fig.~\ref{break_vs_break}. The behavior of the Balmer and 4000 \AA\
  break with age for $\tau_{\rm 10 Myr}$, \td\ and $\tau_{\rm 10 Gyr}$
  (which is comparable to constant star formation for galaxies at
  $z>2$) models are overplotted for the first 3 Gyr. This figure is
  useful for discriminating between different star formation
  histories, as it provides a clear illustration that (near-)constant
  star formation models are not able to produce a large break within 3
  Gyr. However, if a galaxy is extremely dusty, reddening can mimic a
  very large break. All three galaxies fall on the model curves. For
  MS1054-1319 the break implies an age of about 1 Gyr. The break of
  CDFS-695 reveals a younger stellar population. The large Balmer
  break for HDFS-5710 suggests that this galaxy is in a post-starburst
  phase. Unfortunately the errors are too high to draw firm
  conclusions from these diagnostics.

\begin{deluxetable}{cccc}
  \tabletypesize{\footnotesize} 
  \tablecaption{Spectral features\label{diag}}
  \tablewidth{0pt}
  \tablehead{ \colhead{id} & \colhead{\wha(\AA)\tablenotemark{a}} &
    \colhead{$D_n(4000)$} & \colhead{$D_B$}}
  \startdata
  CDFS-695 & 99$\pm$10\tablenotemark{b} &  
  1.17$^{+0.14}_{-0.09}$ & 1.84$^{+0.24}_{-0.15}$ \\
  & &  & \\  
  MS1054-1319 & 19$\pm$4\tablenotemark{c} &
  1.28$^{+0.16}_{-0.04}$ & 1.69$^{+0.24}_{-0.20}$ \\
  & &  & \\
  HDFS-5710 & $<10$ & 1.38$^{+0.47}_{-0.33}$ & 2.40$^{+0.67}_{-0.43}$ \\
  \enddata
  \tablenotetext{a}{Uncorrected for Balmer absorption}
  \tablenotetext{b}{\cite{vd05}}
  \tablenotetext{c}{\cite{vd04}}
\end{deluxetable}

  \subsection{Comparison to \ha\ equivalent width}\label{SFH}

  In addition to the break, the equivalent width of \ha\ (\wha) can
  help discriminate between dusty starbursts and evolved stellar
  populations. As \wha\ is sensitive to the ratio of the current and
  past star formation it is an independent measure of the star
  formation history and the age of a stellar population.

  The \wha\ of MS1054-1319 and CDFS-695 were measured by
  \cite{vd04,vd05}. For HDFS-5710, no \ha\ emission line is detected,
  and thus we give an 3$\sigma$ upper limit for $W_{\rm H\alpha}$
  using the redshift derived in Sect.~\ref{mod}, and assuming a FWHM
  of 500 km/s. To check for a possible \ha\ detection within the
  2$\sigma$ allowed redshift range, we measured the S/N of a possible
  line for redshifts between 2.2 and 2.5 (in steps of 0.001). For none
  of these redshifts do we find a detection exceeding the limits
  derived for the best fitting redshift. The values of \wha\ for all
  galaxies are listed in Table~\ref{diag}. 

\begin{figure*}
  \begin{center}
    \epsscale{1.1}
    \plotone{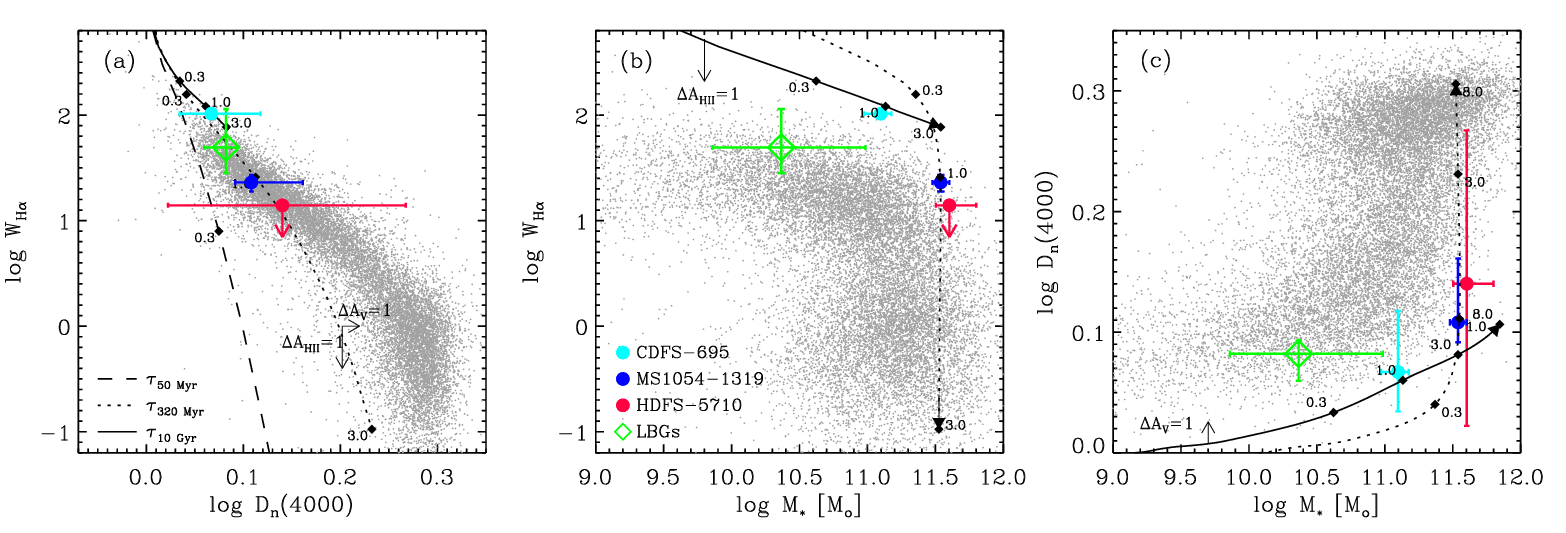}
    \caption{Comparison of spectral properties of the three studied
      galaxies with LBGs \citep[{\em green},][]{er03}, and SDSS
      galaxies \citep[{\em small grey
      dots},][]{tr04,ka03a,ka03b}. Evolution tracks for $\tau_{\rm 10
      Gyr}$ ({\em solid line}), \td ({\em dotted line}) and $\tau_{\rm
      50 Myr}$ ({\em dashed line}) models, derived from the
      \cite{bc03} library and the \cite{ke98} law are overplotted
      without correction for dust. The small black diamonds and the
      corresponding values indicate the age in Gyr. The correction for
      the integrated attenuation of the whole galaxy and extra
      extinction towards HII regions on the models are indicated by
      vectors. (a) \wha\ versus $D_n(4000)$. (b) \wha\ versus the
      stellar mass. (c) $D_n(4000)$ versus stellar mass.  The masses
      of the galaxies are derived from the best-fit stellar
      model. \label{fig_diag}}
  \end{center}
\end{figure*}

  In Fig.~\ref{fig_diag}a we plot \wha\ as a function of $D_n(4000)$
  for the galaxies and different models ($\tau_{\rm 50 Myr}$, \td\ and
  $\tau_{\rm 10 Gyr}$). The plotted \wha\ are corrected for a Balmer
  absorption with an equivalent width of $4 \pm 1$\AA, which is
  typical for stellar populations with ages of 0.5-2.0 Gyr
  \citep{fo04}. The model tracks are derived from the \cite{ke98} law,
  which relates the luminosity of \ha\ to the SFR, in combination with
  the \cite{bc03} models. These tracks illustrate that neither a large
  break nor a low \wha\ can be produced for $\tau_{\rm 10 gyr}$ or CSF
  models within 3 Gyr.

  Fig.~\ref{fig_diag}a illustrates that the three galaxies are at
  different stages in their stellar evolution. The undetected \ha\ of
  HDFS-5710 is consistent with the large break, as both imply an
  evolved stellar population. Furthermore we note that the SFR from
  the continuum fits (see \S \ref{proc}) of 19 M$_{\odot}$/yr is in
  agreement with the SFR upper limit of 25 M$_{\odot}$/yr, derived
  from the 3$\sigma$ upper limit on \ha\ and assuming the best-fit
  \av\ of 0.9 mag.  For MS1054-1319 \wha\ provides additional evidence
  that the stellar light is dominated by evolved stars. And, in
  agreement with all the modeling results, CDFS-695 is still forming
  stars at a high rate.

  We note that for CDFS-695 and MS1054-1319 there are indications that
  \ha\ is not entirely due to photo-ionization from star
  formation. The unusually high [NII]/\ha\ ratios could suggest that
  another source of ionization is contributing to \wha\
  \citep{vd04,vd05} of these two galaxies.

  \subsection{Active Galactic Nuclei?}

  Optical spectroscopy, X-ray studies, and 24$\mu$m imaging infer that
  about 5\% to 25\% of the DRGs show signs of AGN activity
  \citep[Wuyts et al. in prep.;][]{ru04,re05,pa06,we06}, depending on
  $K$-band depth of the sample and the type of AGN. AGNs affect
  rest-frame optical spectra in two ways; they contaminate the stellar
  continuum emission and produce characteristic emission line shapes
  and ratios. Continuum contribution from an AGN weakens the stellar
  break strength, and the strong breaks for HDFS-5710 and MS1054-1319
  (see Fig.~\ref{break_vs_break}) imply that an AGN cannot be the
  dominant contributer to the rest-frame optical continua of these
  galaxies. For CDFS-695 the observed optical break is relatively weak
  and a significant AGN contribution to the continuum cannot be ruled
  out. We note that a minor nuclear contribution, as for a Seyfert 2
  (Fig.~\ref{break_vs_break}), would imply that the stellar breaks are
  even stronger, which would yield older ages for the galaxies.

  There is additional evidence that MS1054-1319 and CDFS-695 do not
  host an AGN which dominates the rest-frame optical continuum. NICMOS
  imaging of MS1054-1319 shows that the emission from this galaxy is
  extended (Toft et al. in prep). Furthermore, this galaxy is not
  detected in X-rays \citep{ru04} and there is no indication for an
  AGN in the rest-frame UV-spectrum \citep[][ID 1671]{vd03}. For
  CDFS-695 the rest-frame UV spectrum lacks AGN features
  \citep[][]{da04a}, although the unusual high [NII]/\ha\ ratios
  mentioned in the previous section could indicate an AGN contribution
  to this galaxy \citep{vd05}.

\section{COMPARISON TO OTHER GALAXIES}

In the previous sections we learned through modeling and measuring the
spectra that the DRG sample contains galaxies with a wide variety of
properties. In this section we compare the measured and modeled
properties with those of other high-z and low-z galaxy samples.

  \subsection{Comparison to other high-redshift galaxies}

  Several studies find that at similar rest-frame V magnitude and
  redshifts, DRGs are older, more massive, and dustier than LBGs
  \citep{vd04,fo04,la05}. We compare in this section the properties of
  the studied DRGs with those of LBGs at similar redshift, using the
  spectral indices and modeling results derived in previous
  sections. Ideally the samples should be matched in $K$ band
  magnitudes or stellar masses. As such matched samples are not
  available caution is required when interpreting the differences.

  As there are no optical break measurements of LBGs, we have to rely
  on the `predicted' break, based on the best fitted SEDs. We used the
  BX and MD galaxies presented by \cite{er03}, which have a median
  redshift of $z\sim2.3$. The photometry and best fit models of the
  galaxies are kindly provided by D.K. Erb; 4 galaxies of these are
  presented by \cite{sh05}. The break has been derived for each of the
  12 galaxies in the Q1623 and Q1700 fields individually and yield a
  median $D_n(4000)$ of 1.21$^{+0.03}_{-0.06}$.  The \wha\ for the
  BX/MD sample is derived from the $L_{\rm H\alpha}$ presented by
  \cite{er03} and the best fit models. These values are quite
  uncertain as slit-losses can introduce errors of up to about 50\% in
  the absolute fluxes \citep{pe01}. The predicted $D_n(4000)$ and
  \wha\ of the BX/MD sample are plotted in Fig.~\ref{fig_diag}a.

  While our galaxies span different stages of stellar population
  evolution, the LBGs are located on the young and star forming part
  of the plot (Fig.~\ref{fig_diag}a). Both DRGs and LBGs fall nicely
  on the predicted \td\ track.

  The LBGs plotted here are all fainter than $K=20$, and thus it could
  be that brighter LBGs have properties more similar to those of the
  plotted DRGs \citep{ad05,sh04}. Also the difference in stellar mass
  of the LBGs and DRGs could be the cause of the different properties
  found for both samples. To show how both properties relate to
  stellar mass, we plotted \wha\ and $D_n(4000)$ versus the stellar
  mass in Fig.~\ref{fig_diag}b and \ref{fig_diag}c respectively. The
  DRGs MS1054-1319 and HDFS-5710 do not only have larger breaks and
  smaller \wha\ than the BX/MD galaxies from \cite{er03}, but have
  also much higher stellar masses. As they span different ranges in
  stellar mass, it is difficult to directly compare the properties of
  the published samples, or interpret it in the context of
  evolutionary scenarios linking LBGs and DRGs.
  
  \subsection{Comparison with low-z galaxies}

  To examine the implications for the evolutionary histories of nearby
  galaxies we now compare the breaks, \wha, and masses of $z>2$ galaxy
  populations with those of SDSS galaxies. The $D_n(4000)$ and \wha\
  (uncorrected for dust) of the SDSS galaxies are adopted from the
  catalogs presented in \cite{ka03a} and \cite{tr04} respectively. We
  used the median dust-corrected stellar masses
  \citep{ka03a}. \cite{ka03a} used the same stellar library
  \citep{bc03} to derive the stellar masses, but adopted the IMF by
  \cite{kr01}. At a given rest-frame $V$-band luminosity, the
  \cite{kr01} IMF yields masses which are about a factor of 2 lower
  than the masses obtained using a \cite{sa55} IMF between 0.1-100
  M$_{\odot}$ \citep{bc03}, and we applied this correction to the SDSS
  galaxies.

  In Fig.~\ref{fig_diag}a the SDSS galaxies show a tight relation
  between $D_n(4000)$ and \wha.  Remarkably, DRGs and LBG fall on this
  same low-redshift relation. The similarity of the $z\sim 2.3$
  galaxies and SDSS galaxies breaks down when we include the stellar
  mass in Figs.~\ref{fig_diag}b and c. At a given mass the $z\sim2.3$
  galaxies have on average smaller breaks and higher \wha\ than the
  low-redshift galaxies. This implies that they are younger and have
  higher specific SFRs than their low-redshift analogs. Furthermore,
  galaxies with the properties observed for the $z\sim2.3$ DRGs are
  rare at low redshift. 

  To show how simplified SFHs for HDFS-5710 and MS1054-1319 behave in
  this plot, we plotted the tracks for the \td\ and $\tau_{\rm 10
  Gyr}$ models, scaled such that both models form the stellar mass of
  MS1054-1319 and HDFS-5710 within 3 Gyrs.  A straightforward
  explanation for the location of DRGs on Fig.~\ref{fig_diag}a--c is
  that they evolve into galaxies with properties comparable to those
  of the most massive and oldest galaxies in the SDSS.  This may imply
  that the DRGs are the younger progenitors of galaxies of similar
  mass today.

\section{SUMMARY AND CONCLUSIONS}

We have presented rest-frame optical spectra of three DRGs at
$z\sim2.3$. To avoid a bias towards galaxies with strong UV emission,
we selected objects which did not necessarily have a spectroscopic
redshift from emission lines prior to the observations. We were able
to measure the continuum shape and directly detect the Balmer/4000
\AA\ break for all galaxies we observed for this purpose. This
presents a significant advance beyond previous studies which were
based on broadband photometry \citep[e.g.][]{pa01,sh01,ru03,fo04,pa06}
or rest-frame optical emission lines \citep[e.g.][]{er03,vd04,sw04}.

We have explored how the direct measurement of the optical continuum
shapes and in particular the Balmer and 4000 \AA\ break can contribute
to our understanding of $z>2$ galaxies. First, we have demonstrated
that fairly accurate redshifts for red $z\sim 2.3$ galaxies can be
obtained by modeling the rest-frame optical continuum around the
Balmer/4000\,\AA\ break. The uncertainties on the modeled redshifts
range from $\Delta z/(1+z)\sim 0.001-0.04$, depending on the S/N of
the spectrum and the strength of the break. Our continuum redshifts
agree very well with the emission line redshifts available for 2 of
the 3 galaxies (within 0.5\%). This opens up the possibility of
deriving redshifts for UV-faint evolved galaxies, which are
underrepresented in high redshift samples selected in optical surveys
and are too faint for optical spectroscopy. Second, we used the
rest-frame optical spectrum to study stellar populations and dust
properties of galaxies, by modeling the spectrum and using spectral
diagnostics. Spectral modeling is very effective in narrowing the
range of possible solutions for age, star formation timescale and dust
content. Including NIR spectra in photometric modeling leads to more
tightly constrained properties of stellar populations in high redshift
galaxies than modeling solely the optical-to-NIR photometry. And
third, the break allows us to constrain the contribution from a
possible AGN. In summary, studying the rest-frame optical shape
appears to be crucial for understanding the red $z>2$ galaxy
population, and provides information which is difficult -- and in some
cases impossible -- to obtain by other means.

The spectral modeling presented here for the three galaxies confirms
the large variation among the SED properties and SFHs of DRGs
\citep[e.g.][]{fo04,la05,we06}, as our objects range from a dusty
starburst with a small break to an apparently evolved galaxy with a
strong 4000 \AA\ break and no detected \ha\ emission. For the galaxy
without emission lines, we derived the redshift from the spectral
continuum shape. The stellar masses of the two most evolved galaxies
(age $\sim$ 1.3-1.4 Gyr) are about $4 \times 10^{11} \rm M_{\odot}$,
and thus they are among the most massive galaxies yet identified at
these redshifts. Comparison to low redshift galaxies suggest that they
probably evolve into galaxies with properties comparable to those of
the most massive and oldest galaxies in the low-redshift Universe. The
younger starburst galaxy (0.3 Gyr) has a stellar mass of $1\times
10^{11} \rm M_{\odot}$.  The strong breaks in the two most evolved
galaxies imply that stellar light is the dominant contributor to the
optical emission. For the younger galaxy a significant contribution
from an AGN cannot be ruled out.

The next step to increase our understanding of high-redshift galaxy
populations is to extend our NIR spectroscopic sample. The present
sample does not allow any statistical analysis, and is mainly an
illustration of the capabilities NIR spectroscopy has to offer. A
larger sample will tell us more about the ratio of evolved versus
star-forming $z>2$ galaxies, and the contribution of AGNs.

\acknowledgments This research would have been impossible without the
generosity and flexibility of the staff of the Gemini Observatory. We
are grateful to the referee for very useful comments, that greatly
improved the paper. We thank Dawn Erb for providing the photometry and
model parameters of the BX/MD sample. This research was supported by
grants from the Netherlands Foundation for Research (NWO), and the
Leids Kerkhoven-Bosscha Fonds. Support from National Science
Foundation grant NSF CAREER AST-0449678 is gratefully acknowledged. DM
is supported by NASA LTSA NNG04GE12G. EG is supported by the National
Science Foundation under Grant No. AST-0201667, an NSF Astronomy and
Astrophysics Postdoctoral Fellowship (AAPF). ST acknowledges the
support of the Danish Natural Research Council. The Keck Observatory
was made possible by the generous financial support of the W.M. Keck
Foundation. The authors wish to recognize and acknowledge the very
significant cultural role and reverence that the summit of Mauna Kea
has always had within the indigenous Hawaiian community.  We are most
fortunate to have the opportunity to conduct observations from this
mountain.

\end{document}